\documentclass{article} % For LaTeX2e
\usepackage{colm}

\usepackage{listings}
\usepackage{microtype}
\usepackage{hyperref}
\usepackage{url}
\usepackage{booktabs}
\usepackage{mathtools}
\definecolor{darkblue}{rgb}{0, 0, 0.5}

\hypersetup{colorlinks=true, citecolor=darkblue, linkcolor=darkblue, urlcolor=darkblue}
\usepackage{array}
\usepackage[pdftex]{graphicx}
\usepackage{diagbox}

\usepackage{subcaption}
\usepackage{arydshln}

\colmfinalcopy

\begin{document}

%% Title
\title{\center Managing multiple agents \\by automatically adjusting incentives}
\author{Shunichi Akatsuka \thanks{Corresponding author, https://orcid.org/0000-0002-1681-6405}\\
Hitachi, Ltd., R\&D Group\\
Kokubunji, Tokyo, Japan \\
\texttt{shunichi.akatsuka.bo@hitachi.com} \\
\And
Yaemi Teramoto \\
Hitachi, Ltd., R\&D Group\\
Kokubunji, Tokyo, Japan \\
\texttt{yaemi.teramoto.xy@hitachi.com} \\
\And
Aaron Courville\\
Mila / Universit\'e de Montr\'eal,\\
Montr\'eal, Canada\\
\texttt{aaron.courville@gmail.com} \\
}

\maketitle
\begin{abstract}
  In the coming years, AI agents will be used for making more complex decisions,
  including in situations involving many different groups of people.
  One big challenge is that AI agent tends to act in its own interest,
  unlike humans who often think about what will be the best for everyone in the long run.
  In this paper, we explore a method to get self-interested agents to work
  towards goals that benefit society as a whole.
  We propose a method to add a manager agent to mediate agent interactions
  by assigning incentives to certain actions.
  We tested our method with a supply-chain management problem and showed that this framework
  (1) increases the raw reward by 22.2\%,
  (2) increases the agents' reward by 23.8\%, and
  (3) increases the manager's reward by 20.1\%.

  \vspace{0.5em}
  \textbf{keywords}: reinforcement learning, general-sum games, multi agent, incentive design, supply chain management 
\end{abstract}

%% Introduction
\section{Introduction}

The rise of AI and machine learning techniques is changing how society operates, as we see more scenarios where AI and machine learning agents play important roles.
From automating routine tasks to enabling sophisticated data analysis, schedule optimization, interactive chatbots, and even robotic manipulations,
AI is reshaping the landscape of work and everyday life.
In the coming years, these AI agents will be used for making more complex decisions, including in situations involving many different groups of people.
One big challenge is that AI agent tends to act in its own interest, unlike humans who often think about what will be the best for everyone in the long run.
This raises the question: How can we get self-interested agent to work towards goals that benefit society as a whole?

Game theory offers a powerful framework to analyze these dynamics.
At its core, the challenge is to foster cooperation among self-interested agents in a scenario that has general-sum payoff structure.
Our goal is to get self-interested agents to cooperate in general-sum games, in a scalable way.
We believe that deep reinforcement learning (RL) is one of the candidates, as it has been shown to reach the human expert level in complex decision-making such as Atari~\citep{mnih2013playing} and Go~\citep{Silver_2016}.
Multi-agent RL methods have been successful in some cooperation games like StarCraft~\citep{alphastar}. 
However, relatively little effort has been made to apply RL in general-sum games.

One recent approach is to this is to make the agent aware that other agents are also learning at the same time.
LOLA~\citep{lola} agent optimizes its policy with the assumption that the other agent is a na\"ive-learning agent.
COLA~\citep{willi2022cola} fixes the consistency problem that LOLA has, and 
POLA~\citep{pola} further develops the method by adding proximal objectives.
These methods work well in simple games like iterated prisoner's dilemma or the Coin Game~\citep{lerer2018maintaining} but have not been shown to scale up to large and complex problems.

Our approach is to modify the game by adding another agent, a manager, which is inspired by the idea of Token Economy.
The objective of the manager is to maximize the sum of the rewards of all agents. 
The manager mediates the agent interactions by showing auxiliary states and giving incentives to take certain actions.
This approach is applicable in the real world when there is an independent organization that benefits from the agents' profits and is able to pay back the profit to the agents.
We can view our method as a type of Automated Mechanism Design~\citep{automatedMechanismDesign}.
However, it differs from previous works in that our approach deals with multi-step optimization problems and that it assumes the agents dynamically learn as the manager learns the policy (mechanism).

%% Method
\section{Proposed Method}
\label{sec:proposal}

\subsection{General Multi-Agent Reinforcement Learning (MARL)}
A reinforcement learning agent is modeled to take sequential actions in an environment formulated as a Markov Decision Process (MDP).
An MDP is defined by the tuple ${<S, A, P, r, \gamma>}$, where $S$ is the state space, $A$ is the action space, $P$ is the transition probability, $r$ is the reward function, and ${\gamma \in [0,1)}$ is the discount factor.
In an MDP environment, an agent observes a state $s_t$ and executes action $a_t$ at timestep $t$.
In the next timestep $t+1$, the environment shifts to a new state $s_{t+1}$ at a probability of ${P = \Pr(s_{t+1} \mid s_t, a_t)}$ and the agent receives a reward ${r(s_t, a_t, s_{t+1})}$.
The goal of the agent is to find a policy ${\pi = \Pr(a_t \mid s_t)}$ that maximizes the discounted total reward $J$, defined by

\begin{equation}  
  J = \Sigma_t \gamma^{t}r(s_t, a_t, s_{t+1}).
  \label{eq:obj_rl}
\end{equation}

In a multi-agent problem, the environment is a multi-agent MDP, which is defined similarly to the MDP with multiple agents taking actions.
In this paper, we focus on a Markov Game~\citep{10.5555/3091574.3091594, zhang2021multiagent}, where all the agents take actions simultaneously at each step.
A Markov Game is defined by the tuple ${<N, S, \{A^i\}, P, \{r^i\}, \gamma>}$ , where $N$ is the number of agents in the environment,
$S$ is the joint state space for all the agents, and $A^i$ is the action space for agent $i$.
The transition function $P$ and the reward function $r^i$ are defined for the joint state and action spaces.
The goal of agent $i$ is to find a policy ${\pi^i=\Pr(a^i_t \mid s^i_t)}$ that maximizes its own discounted reward $J^i$, defined by

\begin{equation}  
  J^i = \Sigma_t \gamma^{t}r^i(s^i_t, a^i_t, s^i_{t+1}).
  \label{eq:obj_marl}
\end{equation}

\subsection{MARL with a manager}
\begin{figure}
    \begin{center}
      %\fbox{
        \includegraphics[width=0.85\hsize, clip, viewport=140 140 660 365]{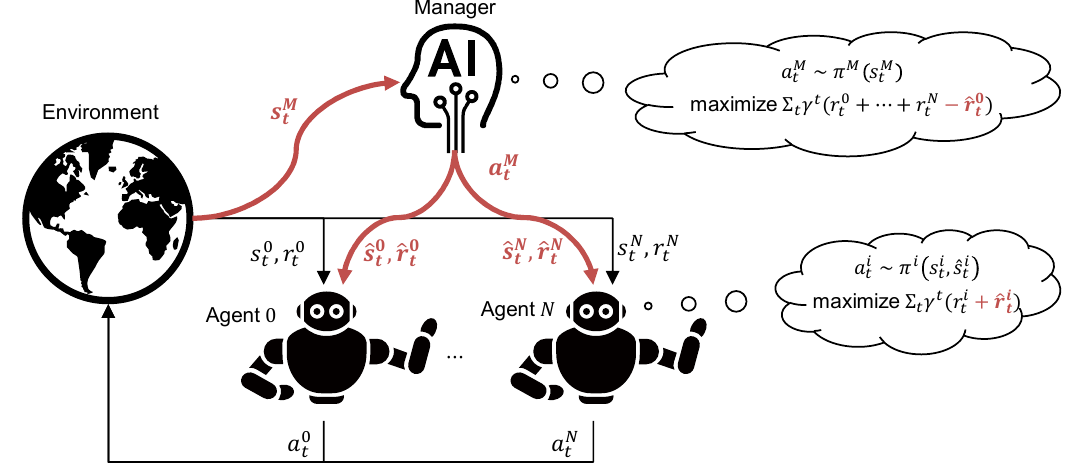}
      %}
      \caption{
        Multi-agent reinforcement learning with a manager.
      }
      \label{fig:managerRL}
    \end{center}
\end{figure}

Figure~\ref{fig:managerRL} shows the overall concept of our method.
In our framework, we add another agent called the manager, which is shown at the top of the figure.
At timestep $t$, the manager observes the state $s^M_t$ and selects an action $a^{M}_t$ according to its policy $\pi^{M}(s^M_t)$.
This action works as an auxiliary state element $\hat{s}^{i}_t$ and supplies auxiliary rewards $\hat{r}^{i}_t$ for agent $i$, such that 
\begin{equation}
  a^{M}_t \vcentcolon= [ \hat{s}^{0}_t, ... , \hat{s}^{N}_t, \hat{r}^{0}_t, ..., \hat{r}^{N}_t ] \sim \pi^{M}(s^M_t),
  \label{eq:act_manager}
\end{equation}
where $N$ is the number of agents.
In other words, the manager tries to control the agents by showing auxiliary states $\hat{s}^{i}_t$ and paying incentives $\hat{r}^{i}_t$ to the agents.
The manager ultimately wants to maximize the sum of the (raw) rewards of all the agents while keeping the paid-out incentive as low as possible,
thus the objective function of the manager $J^{M}$ will become
\begin{equation}
  J^{M} = \Sigma_t \gamma^{t}\{\Sigma_i (r^{i}_t - \hat{r}^{i}_t)\},
  \label{eq:obj_manager}
\end{equation}
where $\gamma$ is the discount factor.

From the agent's point of view, the state and the reward are modified by the manager's action.
For agent $i$, we can convert the state and the reward definition as
\begin{equation}
  \begin{aligned}
    s^{i}_t \leftarrow s^{i}_{t} \oplus \hat{s}^{i}_t,
    \\
    r^{i}_t \leftarrow r^{i}_t + \hat{r}^{i}_t,
  \end{aligned}
  \label{eq:new_state}
\end{equation}
and the objective function for agent $i$, $J^i$, is defined as an ordinary MARL problem defined in Equation~\ref{eq:obj_marl}.

%% Experiments
\section{Experiments}
\label{sec:experiments}

\subsection{The supply-chain optimization problem}
\label{subsec:sco}
\begin{figure}
  \begin{minipage}{0.33\hsize}
    \begin{center}
      %\fbox{
      \includegraphics[width=0.96\hsize, clip, viewport=170 155 550 430]{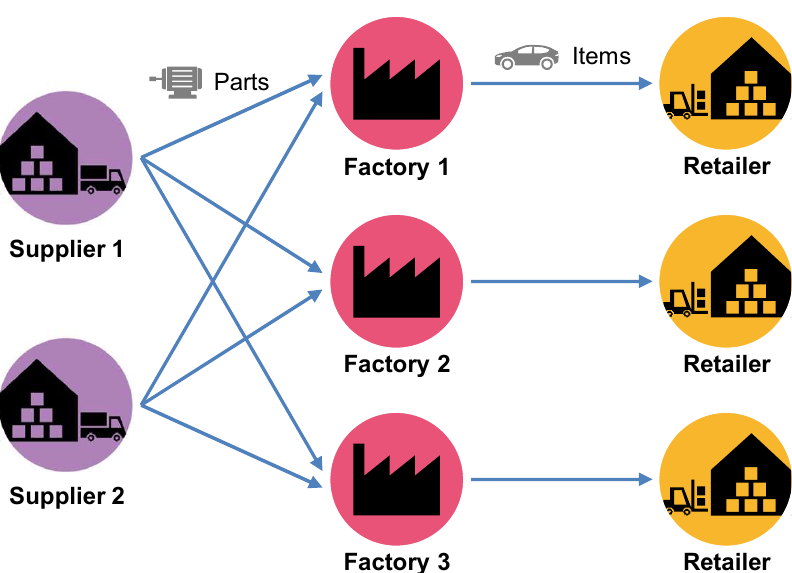}
      %}
      \subcaption{
        Visualization of the supply chain in our experiments.
      }
      \label{fig:supply_chain}
    \end{center}
    \end{minipage}
    \hspace{1mm}
  \begin{minipage}{0.64\hsize}    
    \begin{center}
      %\fbox{
      \includegraphics[width=0.96\hsize, clip, viewport=85 180 670 400]{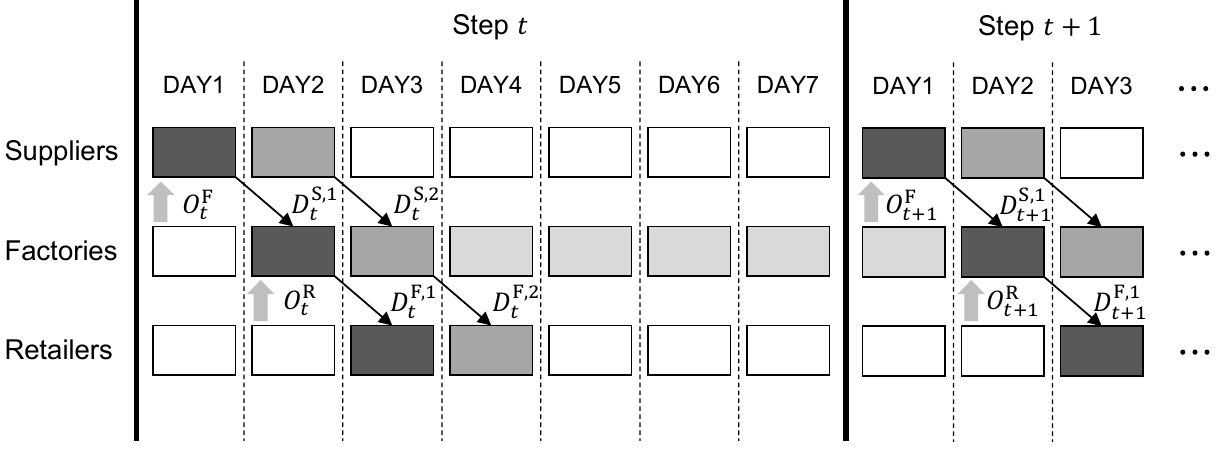}
      %}
      \subcaption{
        Visualization of an environment step. An environment step consists of seven days.
      }
      \label{fig:simulator_timestep}
    \end{center}
    \end{minipage}
    \caption{
        Diagrams showing the supply-chain management environment we used in our experiments.
      }
      \label{fig:simulator}
\end{figure}

We tested our method with a supply-chain optimization problem.
We used a simple supply chain shown in Figure~\ref{fig:supply_chain} with two suppliers, three factories, and three retailers.
The factories can purchase parts from either of the two suppliers, but the retailers can only purchase items from a specific factory.

An environment step consists of seven days, as shown in Figure~\ref{fig:simulator_timestep}.
At the beginning of each step (DAY1), the factories place orders to the suppliers. 
The suppliers produce the parts and deliver them to the factories after several days, depending on the number of orders and the capacity of the suppliers.
The factories assemble the parts to create items, which will take another day.
At every step on DAY2, the factories receive orders from the retailers. 
The factories fulfill the orders by shipping the items to the retailers as early as possible (DAY3$\sim$DAY7), 
and only the items shipped on the earliest possible day (DAY3) are considered as \textit{on time}.
When the number of items that can be shipped is smaller than the number of orders, the remaining orders will be pooled as back orders, which need to be fulfilled in the subsequent steps.
If the number of produced items is larger than the number of orders, then the remaining items will be stored at each factory as stock, which can be used to fulfill future orders.
% The remaining orders/items at the end of a step will be carried over to the next step as back orders/inventories.

The players in this environment are the factories. They decide how many parts to buy from each supplier, to maximize their rewards.
The factory has two objectives, to maximize its \textit{own} profit and to fulfill the retailer's orders on time \textit{as a whole}.
We assume that each supplier has a different price and production capacity per day.
The factories want to buy the parts from cheaper suppliers, but if they all buy from the same supplier,
the number of orders surpasses its capacity and causes delays in parts delivery.
This will decrease the number of items shipped to the retailers on time.
This is where the general-sum-game characteristic shows up
-- some factories should order more from the expensive supplier for timely shipping,
no factory would want to do so as this reduces profit.
% The problem structure is similar to the \textit{tragedy of the commons}.

\subsection{RL formalization}
We set up an RL problem with one agent assigned to each of the three factories.
The action of agent $i$ is to place orders with the two suppliers, thus $a_t^i$ is a two-dimensional integer vector.
We defined the reward of agent $i$ as 
%\begin{equation}
$
  r^i = w^{p} r^{p, i}_t + w^{OFR}r^{OFR}_t, 
$  
%  \label{eq:reward}
%\end{equation}
where $r^{p, i}_t$ is the profit of agent $i$, $r^{OFR}_t$ is the Order Fulfillment Ratio defined in the next paragraph, and $w^p, w^{OFR}$ are the weight factors.

The profit reward is defined as
\begin{equation}
  r^{p, i}  =\frac{1}{C^{p, \textrm{norm}}} \{ N^{\textrm{ship}, i}_t\cdot p^{\textrm{item}} - \Sigma_{s}a_t^{i, s}\cdot p^{\textrm{parts}, s} - I^i_t\cdot p^{\textrm{inventory}} - C^{p, \textrm{offset}}\},
  \label{eq:reward_profit}
\end{equation}
where $N^{\textrm{ship}, i}_t$ is the number of items shipped, $I^i_t$ is the number of inventories at time $t$, $a_t^{i, s}$ is the $s$-th component of the agent action vector $a_t^i$, $p^{\textrm{item}}$ is the selling price of the item, $p^{\textrm{parts}, s} $ is the price of the parts from supplier $s$, and $p^{\textrm{inventory}}$ is the penalty fee imposed to each inventory item.
The index $s$ runs through all the suppliers, i.e. $s=0,1$. The constants $N^{p}$ and $C^{p}$ are the normalization factor and offset parameter, respectively, to normalize the reward range in to approximately $[0,1]$ per step.

The Order Fulfillment Ratio (OFR) is defined by [Number of orders on time]/[Number of total orders]. 
This is an important metric that affects customer satisfaction and efficiency of the supply chain.
We consider a scenario where the OFR is calculated for the supply chain as a whole, and there is a target value $T^{OFR}$ for the OFR.
Then we define the OFR reward $r^{OFR}$ as
\begin{equation}
    r^{OFR}_t =
    \begin{cases}
        1,              & \text{if } \textrm{OFR}\geq T^{OFR}\\
        0,              & \text{otherwise}.
    \end{cases}
    \label{eq:ofr_reward}
\end{equation}

We defined the auxiliary state $\hat{s}^{i}_{t}$ as a two-dimensional vector with values in range $[0, 1]$, and 
the auxiliary reward as $\hat{r}^{i}_t = \hat{s}^{i}_{t-1} \cdot a^{i}_{t-1}$, which is the inner product of the auxiliary state and the agent's action at the previous step.
This means that the incentives are given to factory $i$ for buying the items from suppliers, proportional to the auxiliary state $\hat{s}^{i}_{t}$.
With this definition, the auxiliary reward is calculated from the auxiliary state and thus the manager's action is only to select the auxiliary states,
\begin{equation}
  \hat{a}^{M}_t \vcentcolon= \hat{s}^{0}_{t} \oplus \hat{s}^{1}_{t} \oplus \hat{s}^{2}_{t} \sim \pi^M_t(s^M_t).
  \label{eq:manager_action_exp}
\end{equation}

The raw observation for an agent is a 175-dimension vector, which includes 105 variables on the suppliers' status, 
45 variables on the factory's status, 25 variables on future demand estimates, and 1 variable that indicates the current timestep.
The observation space for the manager is a 531-dimension vector, which includes the observation of all three factories plus a 6-dimensional vector of the agent's actions in the previous step.

The environmental parameters are shown in Table~\ref{tab:parameters}.
We set up the parameters so that supplier~0 is cheaper but has limited capacity,
and supplier~1 is more expensive but has a larger capacity.
The normalization and the offset constants for the profit reward are tuned carefully so that the reward range would be $[0.0, 1.0]$ per step.

\begin{table*}
    \caption{
      List of environmental parameters.
    }
    \begin{center}
    \begin{tabular}{cc}
      \hline
      Capacities per day for suppliers [0, 1] & [100, 450]\\
      $p^{\textrm{item}}, p^{\textrm{parts}, s}, p^{\textrm{inventory}}$ & $10, [2, 5], 1$ \\
      $t_{\textrm{max}}$ & 52 ($\sim$ 1 year) \\
      Retailer's demand & $100 \pm 50$ flat distribution \\
      Demand forecast & demand $+ N(0,30)$\\
      $C^{p, \textrm{norm}}, C^{p, \textrm{offset}}, T^{OFR}$ & 300, 400, 0.8\\
      $w^{p}, w^{OFR}$ & 0.5, 0.5 \\
      \hline
    \end{tabular}
    \label{tab:parameters}
  \end{center}
\end{table*}

\subsection{Training and Evaluation}
\label{subsec:training}
Both the agents and the manager are trained with DDPG~\citep{ddpg}.
The actor and the critic networks for the agents and the manager have two fully-connected hidden layers with 128 nodes each.
The outputs of the actor networks are converted to the range $[0,1]$ with a $\tanh$ function. 
The agents' actions are further converted to integers in the range $[0, 99]$ by multiplying the output by 100 and rounding down.
We train the agents with two frameworks: a na\"ive MARL, where there is no manager, and our proposed framework with the manager.
In both cases, we train the agents and the manager with 500 episodes and with 10 different random seeds.

We take the final 25 episodes of the training to evaluate the performance.
We evaluate the mean scores of the agents and the manager, as well as their standard deviations.

%% Results
\section{Results}
\begin{figure*}
. \begin{minipage}{0.49\hsize}
   \begin{center}
     \includegraphics[width=0.99\hsize]{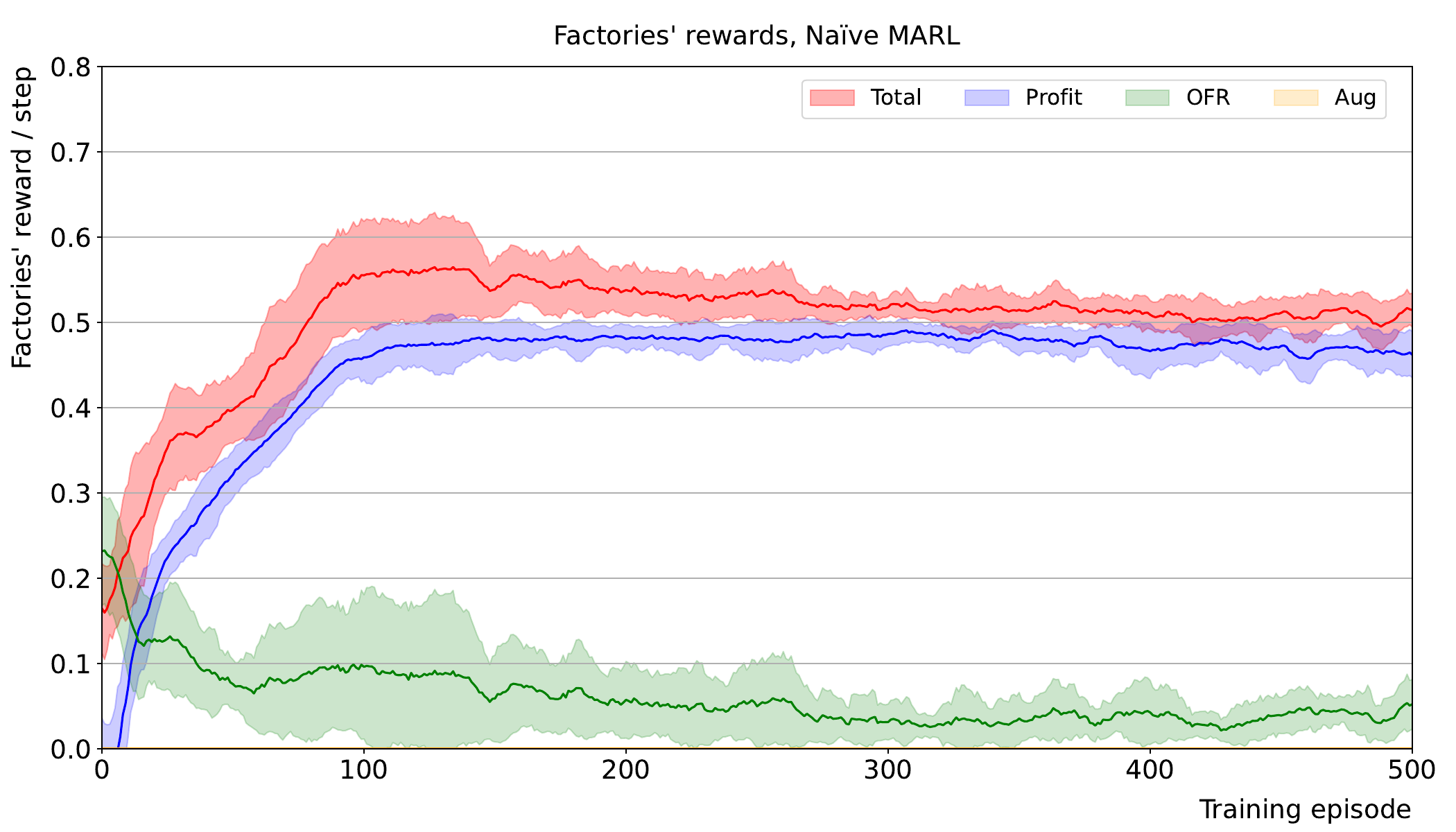}
     \subcaption{
       Na\"ive
     }
     \label{fig:naive_reward}
   \end{center}
 \end{minipage}
 \begin{minipage}{0.49\hsize}
   \begin{center}
     \includegraphics[width=0.99\hsize]{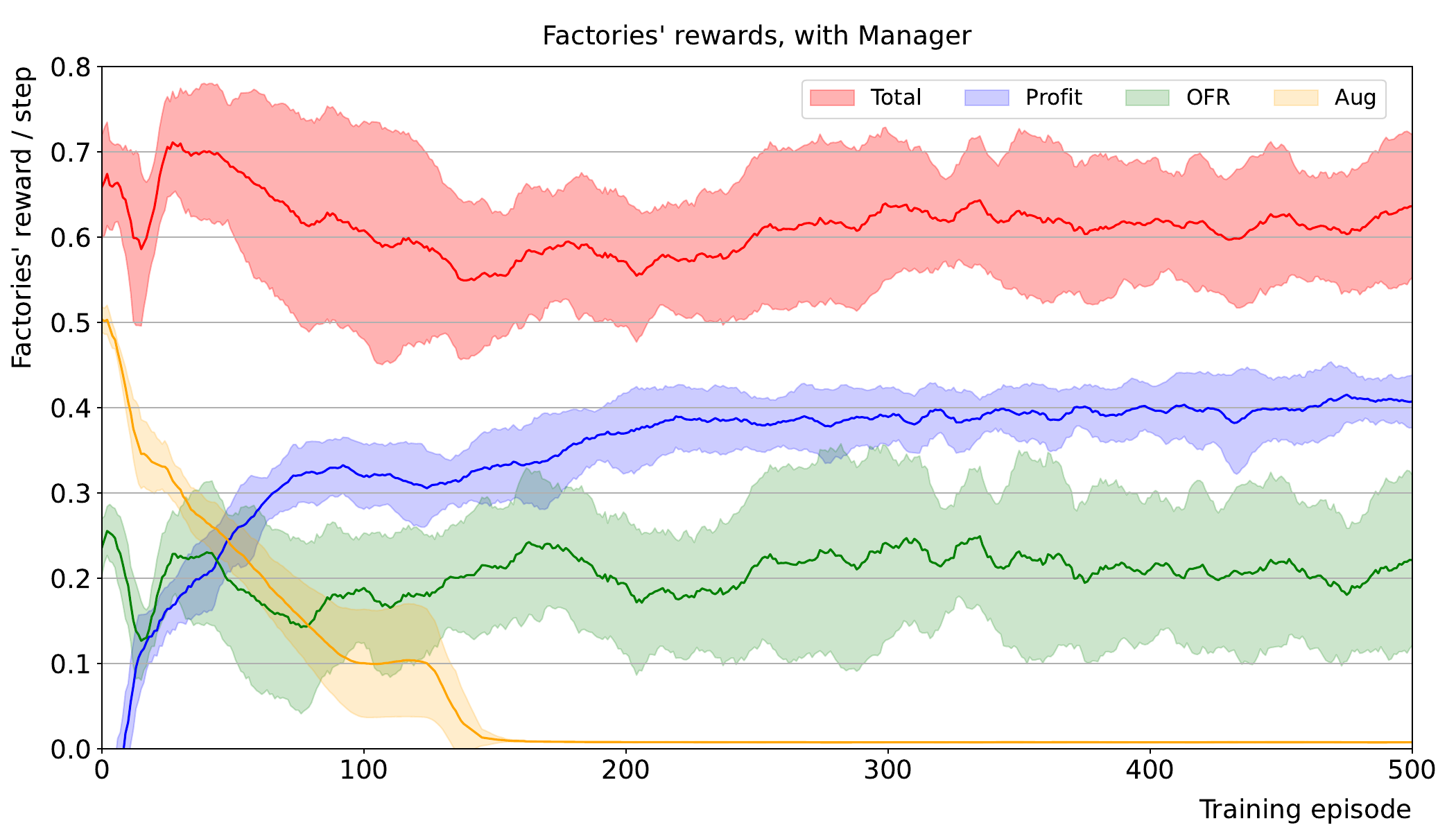}
     \subcaption{
       With Manager
     }
     \label{fig:manager_reward}
   \end{center}
 \end{minipage}
 \caption{
   The average reward of the factories. 
   The error bands indicate the standard deviations of 10 runs with different random seeds.
 }
 \label{fig:reward_summary}
\end{figure*}

Figure~\ref{fig:reward_summary} shows the average reward of the factories during the training.
Figure~\ref{fig:naive_reward} shows the plot for the setup without the manager taking action, and
Figure~\ref{fig:manager_reward} shows the plot with the manager.
These plots show that, with the manager, the profit reward shown in blue decreases but the OFR reward increases more than that,
which improves the final score, shown in red.
Note that the decreasing score at the beginning of the training in Figure~\ref{fig:manager_reward} is caused by
the manager randomly incentivizing the factories, and quickly decreasing the incentives.

\begin{table}[t]
  \begin{center}
    \caption{
      Mean scores of the last 25 episodes in 10 training runs.
      The column ``Factories'' shows the average score for the three factories.
      The row ``Rewards (raw)'' shows the scores without the auxiliary rewards.
    }
    \begin{tabular}{|l|c|c|c|c|c|}
      \hline
      Method & Factories & Manager \\
      \hline \hline
      Na\"ive  & $0.505 \pm 0.025$ & $0.505 \pm 0.025$ \\
      \hline
      w/ Manager & $0.625 \pm 0.079$ & $0.610 \pm 0.079$ \\
      \hdashline
      Rewards (raw) & $0.617 \pm 0.079$ & $0.617 \pm 0.079$ \\
      \hline
    \end{tabular}
    \label{tab:summary}
  \end{center}
\end{table}

The mean scores of the last 25 training episodes are shown in Table~\ref{tab:summary}. Without the manager, the average reward of the factories was $0.505$, and with the manager, it improved to $0.625$, which is a $23.8\%$ improvement. The manager's reward increased to $0.610 \pm 0.051$, which is a $20.1\%$ improvement.
The raw rewards without considering the auxiliary rewards improved to $0.617$, which is a $ 22.2\%$ improvement on average.

\begin{figure*}
  \begin{minipage}{0.49\hsize}
    \begin{center}
      \includegraphics[width=0.99\hsize]{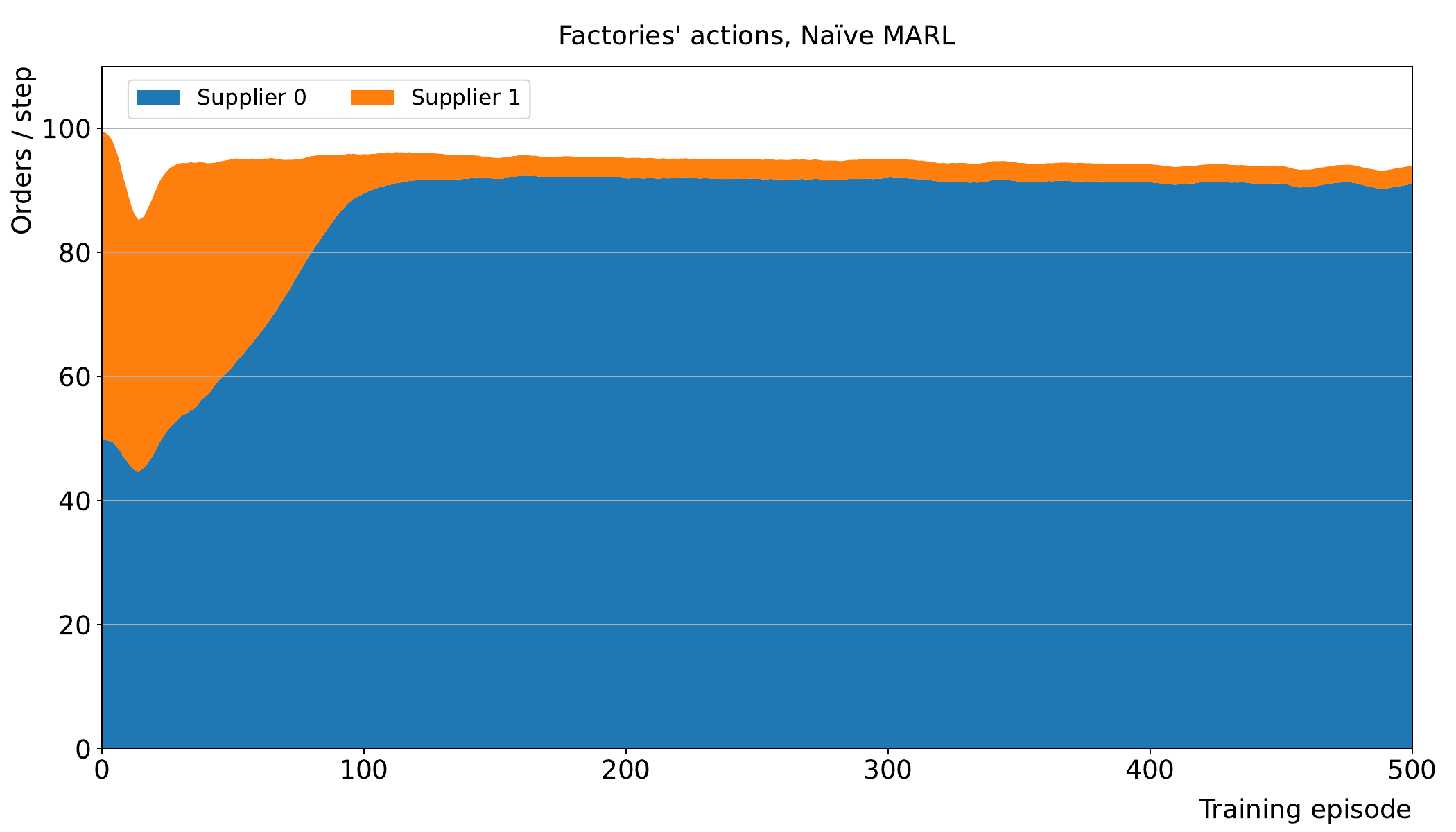}
      \subcaption{
        Na\"ive
      }
      \label{fig:naive_actions}
    \end{center}
  \end{minipage}
  \begin{minipage}{0.49\hsize}
    \begin{center}
      \includegraphics[width=0.99\hsize]{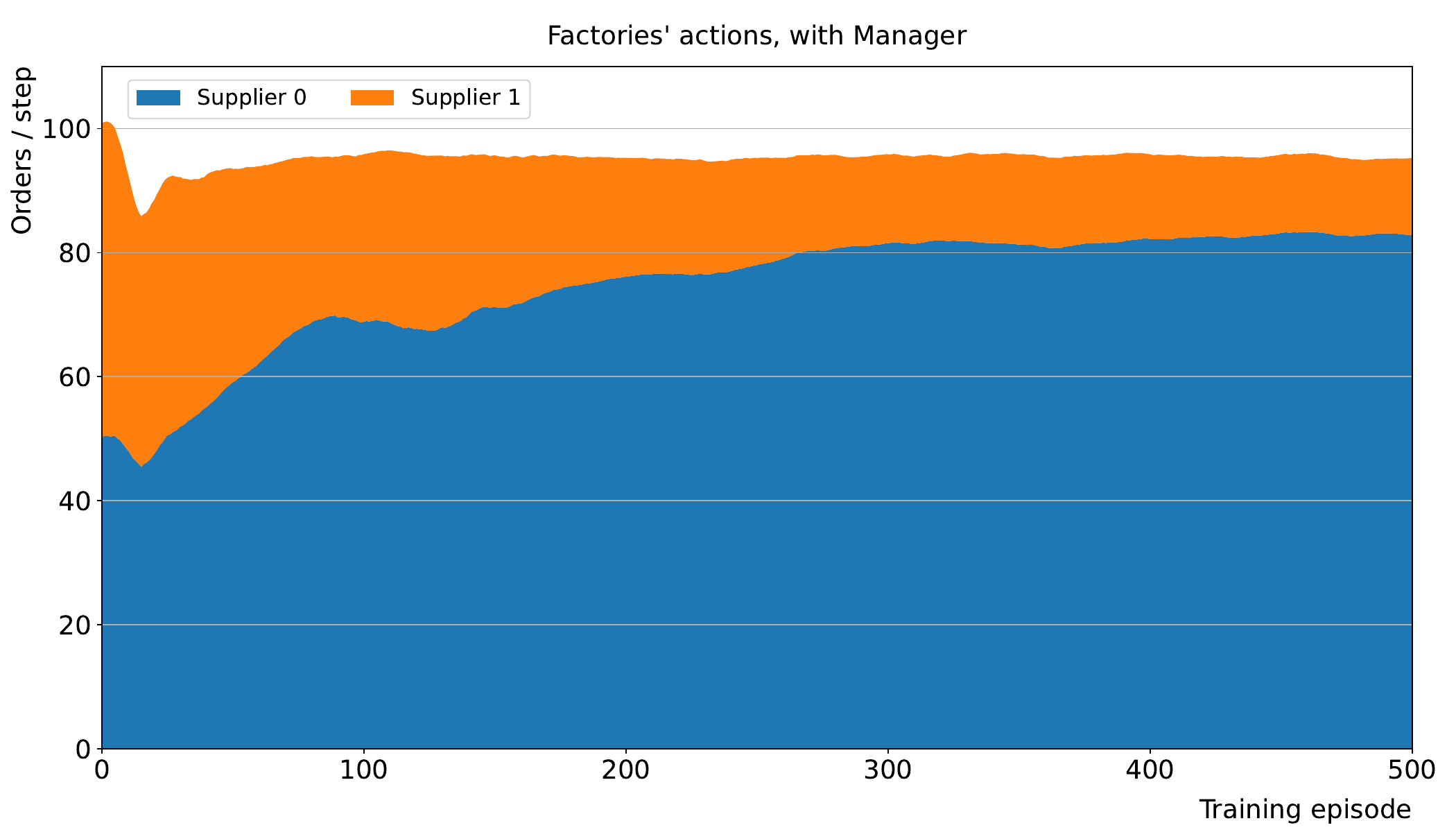}
      \subcaption{
        With Manager
      }
      \label{fig:manager_actions}
    \end{center}
  \end{minipage}
  \caption{
    The actions of the factories without and with the manager.    
  }
  \label{fig:actions}
\end{figure*}

We further investigate how adding the manager influences the factories' actions.
Figures~\ref{fig:naive_actions}, \ref{fig:manager_actions} show the average factory's actions during the training
without and with the manager.
Without the manager, the factories mostly buy from supplier 0, but after adding the manager, the factories buy more from supplier 1. This contributes to improving the OFR significantly.

%% Discussions
\section{Conclusion and Discussion}
In this paper, we tackled the problem of making self-interested agents cooperate in a multi-agent general-sum game environment.
We proposed a method to add an agent called a manager in a multi-agent environment
to mediate agent interactions by assigning incentives to certain actions.
We tested our method with a supply-chain management problem and showed that it increases the total profit of all the players.

Although our experiments show that the manager's policy can be obtained with a simple reinforcement learning framework,
one limitation is that we assume the agents are na\"ive learning RL agents.
This is a strong constraint, because in reality, these agents may be more clever and try to exploit the manager,
or sometimes can be less reasonable and stick to their original policy. % -- or in the extreme, it can even be a human!
It is important to consider how the agent architectures impact the performance of our method.

\section*{Data Availability Statement}
The data that support the findings of this study are available on request from the authors.
Requests for data should be addressed to [Author Name] at [Author's Email Address].

\section*{Conflict of Inetest Statement}
On behalf of all authors, the corresponding author states that there is no conflict of interest.

%% Bib
\bibliography{bibliography}
\bibliographystyle{colm}

\end{document}